# Translation of Fetal Brain Ultrasound Images into Pseudo-MRI Images using Artificial Intelligence

**Naomi Silverstein**
Technion Institute of Technology
Haifa
naomish@campus.technion.ac.il

**Efrat Leibowitz**
Rambam Health Care Campus
Haifa
e_leibowitz@rambam.health.gov.il

**Ron Beloosesky**
Rambam Health Care Campus
Haifa
ronbel3@gmail.com

**Haim Azhari**
Technion Institute of Technology
Haifa
haim@bm.technion.ac.il

## Abstract

Ultrasound is a widely accessible and cost-effective medical imaging tool commonly used for prenatal evaluation of the fetal brain. However, it has limitations, particularly in the third trimester, where the complexity of the fetal brain requires high image quality for extracting quantitative data. In contrast, magnetic resonance imaging (MRI) offers superior image quality and tissue differentiation but is less available, expensive, and requires time-consuming acquisition. Thus, transforming ultrasonic images into an MRI-mimicking display may be advantageous and allow better tissue anatomy presentation. To address this goal, we have examined the use of artificial intelligence, implementing a diffusion model renowned for generating high-quality images. The proposed method, termed "Dual Diffusion Imposed Correlation" (DDIC), leverages a diffusion-based translation methodology, assuming a shared latent space between ultrasound and MRI domains. Model training was obtained utilizing the "HC18" dataset for ultrasound and the "CRL fetal brain atlas" along with the "FeTA " datasets for MRI. The generated pseudo-MRI images provide notable improvements in visual discrimination of brain tissue, especially in the lateral ventricles and the Sylvian fissure, characterized by enhanced contrast clarity. Improvement was demonstrated in Mutual information, Peak signal-to-noise ratio, Fréchet Inception Distance, and Contrast-to-noise ratio. Findings from these evaluations indicate statistically significant superior performance of the DDIC compared to other translation methodologies. In addition, a Medical Opinion Test was obtained from 5 gynecologists. The results demonstrated display improvement in 81% of the tested images. In conclusion, the presented pseudo-MRI images hold the potential for streamlining diagnosis and enhancing clinical outcomes through improved representation.

***Keywords*** Diffusion Model, Deep-Learning, Image Translation, Ultrasound, MRI, Fetal, Lateral-Ventricles, Sylvian-Fissure, Correlation.

# 1 Introduction

Ultrasound (US) stands out as a highly cost-effective modality within contemporary medical imaging. Its non-invasive attributes, prevalent availability, and safety make it applicable across diverse medical domains. Despite its pronounced merits, the US has notable disadvantages. Particularly, ultrasound-generated images exhibit a suboptimal signal-to-noise ratio (SNR). Moreover, the images are characterized by considerable speckle noise, resulting in limited anatomical visualization. As a result, clinical insights are compromised and necessitate a substantial reliance on the interpretative skills of the radiologist (1).

Conversely, magnetic resonance imaging (MRI) offers enhanced soft tissue differentiation, a significantly higher Signal-to-Noise Ratio (SNR), and better anatomical display. However, the trade-off lies in the protracted duration required for MRI image acquisition, the need for substantial infrastructure, high cost, and limited accessibility compared to ultrasound (2). Consequently, the potential transformation of ultrasonic images into a display akin to that of MRI promises potentially improved diagnostic capabilities.

In recent years, neural networks have made significant progress, especially in the field of image generation and image translation. One of the first architectures to achieve the image translation goal was the Cycle Generative Adversarial Network (CycleGAN) (3), followed by more GAN-based architectures (4–6). Within the domain of US-MRI translation, several studies have been conducted employing GAN-based methodologies for US-to-MRI conversion (7–9). Similarly, there have been significant advances in iterative generative models, such as score-based models (10) and diffusion models (11,12), which have exhibited the capacity to generate synthetic images of superior quality comparable to those produced by contemporary GAN methodologies (13). These diffusion models operate by systematically adding Gaussian noise to an image until the image reaches a state of complete white noise. Subsequently, a model is trained to execute the inverse process, gradually diminishing Gaussian noise from the white noise until attaining a coherent, high-quality image. In the field of diffusion-based image translation, contrary to GAN-based architectures, relevant studies are absent (to the best of our knowledge), addressing the synthesis of MRI images from US data.

An important component of routine prenatal ultrasound anomaly scans is the assessment of the posterior horns of the fetal lateral cerebral ventricles (14). The International Society of Ultrasound in Obstetrics and Gynecology (ISUOG) guidelines note that: "due to artifacts in the near field of the image, caused by shadowing from the proximal parietal bone, in the standard trans ventricular plane, only the hemisphere and the lateral ventricle on the far side of the transducer are usually visualized clearly " (15). However, evaluating both ventricles is clinically significant, as fetal cerebral lateral ventriculomegaly is among the most common findings in routine prenatal ultrasounds (16). Additionally, the ability to identify the shape of the Sylvian fissure is crucial during routine ultrasound examinations, as it can serve as an indicator of cortical dysplasia. The Sylvian fissure is typically more distinctly visible on both sides on MRI images (17).

It is postulated here that there may be a potential clinical benefit in transforming US images into pseudo-MRI images, thereby enabling the incorporation of the high-quality visibility characteristic in MRI images into a cost-effective and expeditious ultrasound machine. The hypothesis advanced here suggests that such a transformation can be realized through the implementation of a modified diffusion-based translation model. The modified model suggested here imposes correlation between the source and outcome images in every step during the reconstruction process to better preserve the anatomical information and is therefore titled Dual Diffusion Imposed Correlation (DDIC).

# 2 Methods

## 2.1 Diffusion models

Initially, we describe the training methodology of diffusion models as introduced by Ho *et al.* (11) "Denoising Diffusion Probabilistic Models" (DDPM). Given an image $x_0$ drawn from a source data distribution $q(x_0)$, the forward process is established by the addition of a small amount of Gaussian noise to $x_0$ over T steps, with step sizes governed by the variance schedule $\{\beta_t \in (0, 1)\}_{t=1}^{T}$. At each step, given the image $x_{t-1}$, the resulting next image $x_t$ is derived from the incorporation of Gaussian noise. The definition of a forward step $q(x_t \mid x_{t-1})$ is articulated as follows:

$$q(x_t \mid x_{t-1}) = N(x_t; \sqrt{1-\beta_t} \cdot x_{t-1}, \beta_t I).$$

Eq. (1)

Defining $\alpha_t = 1 - \beta_t$ and $\bar{\alpha}_t = \prod_{i=1}^{t} \alpha_i$, the final $x_T$ can be calculated in a single step:

$$x_T = \sqrt{\bar{\alpha}_T} \cdot x_0 + \sqrt{1-\bar{\alpha}_T} \cdot \epsilon, \quad \text{where} \quad \epsilon \sim \mathcal{N}(0, I).$$

Eq. (2)

To generate an image from the initial condition $p(x_T) \sim \mathcal{N}(x_T; 0, I)$, a necessity arises for the reverse diffusion process. Since the elimination of the noise depends on the distribution of the source data, it becomes necessary to train a model to predict noise removal. The backward process defined by the function $p(x_{t-1} \mid x_t)$, given the noised image $x_t$, the function $p$ calculates the resultant denoised image $x_{t-1}$:

$$p(x_{t-1} \mid x_t) = \mathcal{N}(x_{t-1}; \mu_\theta(x_t, t), \sigma_t^2 I),$$

Eq. (3)

where $\sigma_t^2 = \beta_t$, and $\mu_\theta$ is the average of the added Gaussian noise which the model learns to predict. $\mu_\theta$ is given by:

$$\mu_\theta(x_t, t) = \frac{1}{\sqrt{\alpha_t}} \left( x_t - \frac{1-\alpha_t}{\sqrt{1-\bar{\alpha}_t}} \epsilon_\theta(x_t, t) \right),$$

Eq. (4)

where $\epsilon_\theta(x_t, t)$ is the trained model by minimizing the loss function $L_t$:

$$L_t = \mathbb{E}_{t \sim [1,T], x_0, \epsilon_t} \left[ \left\| \epsilon_t - \epsilon_\theta(\sqrt{\bar{\alpha}_t} \cdot x_0 + \sqrt{1-\bar{\alpha}_t} \cdot \epsilon_t, t) \right\|^2 \right].$$

Eq. (5)

The DDPM sampling method is not deterministic, meaning that different images can result from the same latent noise sample. Thus, we utilize the deterministic sampling method outlined in Denoising Diffusion Implicit Models (DDIM) (18):

$$x_{t+\Delta t} = \sqrt{\frac{\alpha_{t+\Delta t}}{\alpha_t}} \cdot x_t + \left( \sqrt{1-\alpha_{t+\Delta t}} - \sqrt{1-\alpha_t} \cdot \sqrt{\frac{\alpha_{t+\Delta t}}{\alpha_t}} \right) \cdot \epsilon_\theta(x_t, t),$$

Eq. (6)

where $x_t$ is a noised image at timestep t, and $x_{t+\Delta t}$ is the obtained noised or denoised image when $\Delta t$ is positive or negative accordingly.

For conciseness, we define the symbol $ForwardODE^{US}$ for the noising direction, meaning $\Delta t = 1$ and $\epsilon_\theta(x_t, t)$ is the learned denoiser from the US dataset.

$$x_{t+1} = ForwardODE^{US}(x_t, t),$$

Eq. (7)

where $x_t$ designates here the generated US image at timepoint t. In addition, we define the symbols $BackwardODE^{MRI}$ and $BackwardODE^{US}$ for the denoising process, meaning $\Delta t = -1$, and $\epsilon_\theta(x_t, t)$ is the learned MRI dataset denoiser for the former and the learned US dataset denoiser for the later.

$$x_{t-1} = BackwardODE^{US}(x_t, t),$$

Eq. (8)

$$y_{t-1} = BackwardODE^{MRI}(y_t, t),$$

Eq. (9)

where $y_t$ designates the generated MRI image at timepoint t.

Dual Diffusion Implicit Bridge (DDIB) (19) is a method suggested for translating between unpaired images. DDIB comprises two separate diffusion processes, one for each domain. The algorithm involves utilizing $ForwardODE^{domain1}$ to noise the first domain image into a latent code and then denoise the latent into the second domain image using $BackwardODE^{domain2}$.

This method proves efficacious in scenarios where translation is required between objects sharing fundamental structural characteristics and global features, but fails in the translation of objects with finer details. However, in the context of medical imaging translation, wherein even minor details hold significance for clinical diagnosis, the preservation of the precise composition of the source image becomes paramount. Hence, we suggest the Dual Diffusion Imposed Correlation method (DDIC), in which the backward process of the DDIB method mandates correlation between the restored two domains as schematically shown in Figure 1.

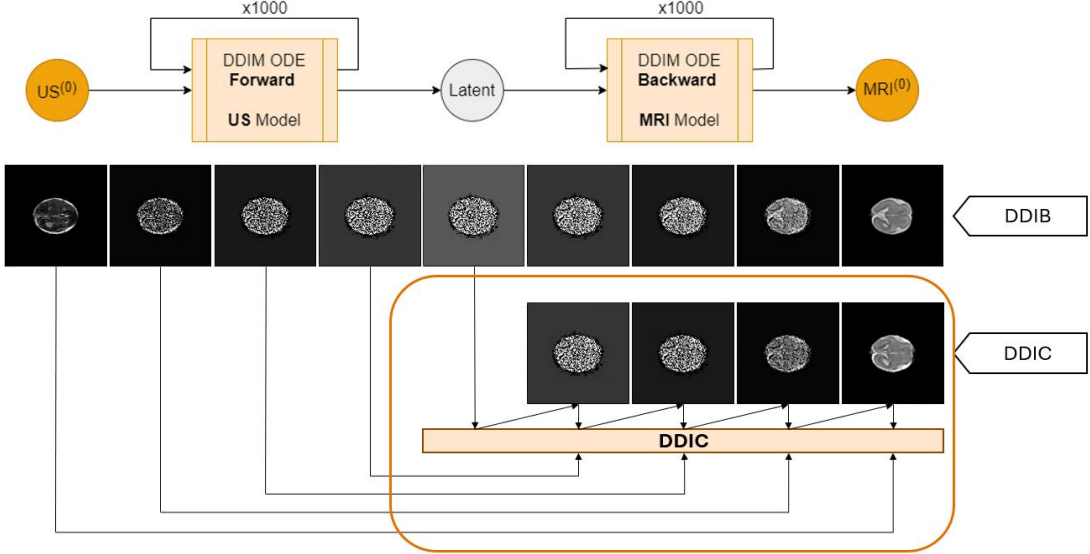

Figure 1: **Top** schematic depiction of the DDIB algorithm diagram and exemplary test sample. **Bottom** schematic depiction of the DDIC. The DDIC diagram is shown in detail in Figure 2.

## 2.2 The DDIC model

Initially, the US image $X$ undergoes denoising through $T$ steps with $ForwardODE^{US}$ utilization, resulting in the extraction of the US latent code. The reconstruction of the pseudo-MRI image from the US latent code is achieved gradually. In each timestep $t$, the model uses $BackwardODE^{MRI}$ to get primary denoised MRI image and use $BackwardODE^{US}$ from the same latent code to get the corresponding parallel US reconstruction. Subsequently, both images undergo a median filtering process, in order to overcome the speckle noise. Next, the correlation between the two median filtered images is calculated, receiving the loss function:

$$loss = -1 \cdot CorrCoef(\tilde{X}_{t-1}, \tilde{Y}_{t-1}),$$

Eq. (10)

where $\tilde{X}_{t-1}, \tilde{Y}_{t-1}$ are the reconstructed US and MRI images after passing the median filter respectively, and the correlation $CorrCoef$ is given by:

$$CorrCoef(a, b) = \frac{cov(a, b)}{\sqrt{cov(a, a) \cdot cov(b, b)}},$$

Eq. (11)

where $cov(a, b)$ is the covariance between images $a$ and $b$.

Our conjecture posits that the efficacy of this approach arises from the median filter's inherent smoothing effect, which enables correlation analysis to emphasize crucial image features rather than being influenced by the intrinsic noise characteristic of the ultrasound imagery.

Following, optimization of $Y_t$, the initial MRI image for the $BackwardODE^{MRI}$, is conducted by calculating the derivative of the loss function according to $Y_t$, adopting the methodology outlined in detail in Parmar et al. (20), which has demonstrated efficacy with maximum cross-attention optimization. This derivation is accomplished via a single-step gradient descent employing an optimal step size selected based on the most favorable outcomes observed during testing. The reconstruction overall consists of $T = 1000$ time steps, the algorithm for each step is illustrated in Figure 2.

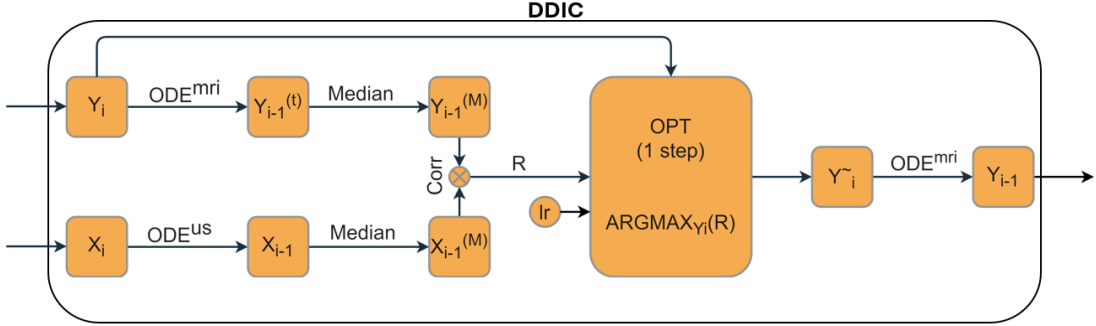

Figure 2: A block diagram of a single DDIC step.

## 2.3 Datasets

The chosen two-dimensional (2D) anatomical cross-section for the US-MRI translation in this study was the "transthalamic" axial plane. This plane serves as a key monitoring tool for assessing fetal development through pregnancy, hence, it is of high clinical importance (21). The gestational age range was chosen to be within 21-38 weeks due to the common overlap between US and MRI examinations during these weeks.

The US images were sourced from the "Automated measurement of fetal Head Circumference" (HC18) dataset (22), encompassing 1000 training images with dimensions of $800 \times 540$ pixels, with a pixel size range of 0.049-0.393 mm². The training set also includes an image with the manual annotation of the head circumference (HC) and its measurement in millimeters for each HC, which was outlined by a trained sonographer. To adhere to the designated gestational age range, only images with HC falling within the range of 170 mm to 350 mm were selected from the training set (23). After excluding images that were either outside the above range or unclear, the final refined US dataset comprised 365 images.

The MRI data employed in this study were downloaded from two distinct datasets: the "CRL fetal brain atlas" database (24) and the "FeTA" challenge dataset (25). The former consists of 18 three-dimensional (3D) T2-weighted scans, with dimensions of $135 \times 189 \times 155$ pixels, pixel size of 0.8 mm³, and covering a gestational age range of 21 to 38 weeks. The latter encompassed 80 3D T2-weighted scans, with dimensions of $256 \times 256 \times 256$ pixels, pixel size of ~0.5 mm³, and gestational age spanning from 20 to 35 weeks. In both datasets, several slices around and including the transthalamic axial plane were extracted, resulting in a combined MRI dataset composed of 251 images.

Due to limitations in computing power, as a proof of concept, all images (US and MRI) used in this study were preprocessed to yield a size of $128 \times 128$ pixels with a pixel size of 1.094 mm². Fetal heads were segmented based on the skull circumference in the US scans and brain tissue in the MRI scans. All images were centered and rotated so that the third ventricle is horizontal in the frame and the lateral ventricles are on the left side.

## 2.4 Evaluation protocols

Importantly, it should be noted that ground truth fetal US-MR exactly paired data are unavailable, as the images from those modalities are not perfectly aligned. This is due to the inability to perform ultrasound and MRI scans simultaneously in the womb. Even when scans are conducted within a short timeframe, fetal movement often results in positional changes, leading to misalignment between slices from the two imaging modalities (26). Therefore, we compared the US images to the pseudo-MRI images using different approaches.

The comparison of image interpretability is inherently subjective, and conventional evaluation metrics, such as Structural Similarity (SSIM), commonly employed in image reconstruction quality analysis, become irrelevant in the context of unpaired US-MRI data. Given this circumstance, we chose the adoption of alternative metrics, specifically the Mutual Information (MI) metric, the Peak Signal-to-Noise Ratio (PSNR) metric, and the Fréchet Inception Distance (FID) for the quantitative evaluation in our study.

MI serves as a metric for image alignment, obviating the necessity for identical signal characteristics in the compared images. It quantifies the predictive capacity regarding the signal in the second image based on the signal intensity observed in the first, thereby offering an assessment of the degree of concordance between the two images. The MI is defined as(27):

$$I(X;Y) = \sum_{y \in Y} \sum_{x \in X} P_{(X,Y)}(x,y) \log\left(\frac{P_{(X,Y)}(x,y)}{P_X(x)P_Y(y)}\right),$$

Eq. (12)

where $P_X$ and $P_Y$ are the marginal probability mass function of X and Y respectively, and $P_{(X,Y)}$ is the joint probability mass function of X and Y.

The PSNR is defined as (28):

$$PSNR = 10 \cdot \log_{10}\left(\frac{MAX_I^2}{MSE}\right),$$

Eq. (13)

where $MAX_I$ is the maximum possible pixel value of the image, 255 in our case, and MSE is the mean square error between the two images.

The FID is a metric employed to evaluate the quality of images generated by a generative model. It compares the distribution of generated synthetic images to that of a set of real images. A lower FID score indicates a greater similarity between the two distributions, with a score of zero signifying identical distributions. In this work, we employed the FID between the set of the generated pseudo-MRI to the set of the source US images (29).

Another evaluation performed is the Contrast-to-Noise Ratio (CNR). This metric was taken to assess the improved contrast of the generated MRI images relative to the original US images. The CNR is defined as:(1)

$$CNR = \frac{|\bar{I}_{ROI} - \bar{I}_{background}|}{\sqrt{\sigma_{ROI}^2 + \sigma_{background}^2}}$$

Eq. (14)

Where $\bar{I}_{ROI}$ and $\bar{I}_{background}$ are the average intensity in a region of interest (ROI) and its background respectively, and $\sigma_{ROI}^2$ and $\sigma_{background}^2$ are their standard deviation. The ROI selected for this evaluation is the distal lateral ventricle.

In addition to these quantitative methods, semi-qualitative evaluation was applied. For that aim, we have utilized the open-source "Segment Anything" algorithm developed by Meta AI (30). The algorithm was applied to the test set images, where the lateral ventricles and the Sylvian fissure were segmented. The segmentation process aims to demonstrate the practical benefits of converting US images into MRI images. It was done on both full-resolution and low-resolution US images and the DDIC-generated pseudo-MRI. The quality of segmentation was visually examined.

Finally, we assessed the clinical benefits of translating images from US to pseudo-MRI qualitatively using the Medical Opinion Test (MOT). A user study was conducted to evaluate MOT performance. Five physicians were each given a set of 40 pairs of US images and their corresponding translated pseudo-MRI images. For each pair, the physicians were asked to indicate whether the translated image provided an advantage in terms of clinical diagnosis based on several parameters. These parameters included: the proximal lateral ventricle, the distal lateral ventricle, the proximal Sylvian fissure, and the distal Sylvian fissure.

## 2.5 Training parameters

The US dataset was divided into a training set, comprising 90% of the data, and a test set, comprising of the remaining 10%. The training set for the MRI model was the complete MRI dataset. The network architecture was based on the framework provided by DDIB with the addition of the DDIC module. The diffusion model hyperparameters were set to timestep 1000, cosine beta scheduler, and $\epsilon_t$-prediction. For the DDIC reconstruction, the gradient step size was set to $lr = 3$ (see Figure 2). The model was trained on an NVIDIA RTX 2080 GPU running under Linux. Training computation time was approximately 48 hours for each model.

# 3 Results

To evaluate the performance of the DDIC model, we have used the 10% test images taken from the US dataset to synthesize pseudo-MRI images. Additionally, DDIB and CycleGAN, were also applied to generate pseudo-MRI images, providing reference outputs for comparative analysis. Figure 3 illustrates a collection of representative images transformed from US to pseudo-MRI by the three methods. Notably, the pseudo-MRI exhibits enhanced visual clarity regarding brain structures, characterized by sharper delineation of borders and improved contrast. It is noteworthy that the algorithm effectively mitigates acoustic shadow artifacts, commonly appearing in ultrasonic images, especially when evaluating the size of the proximal lateral ventricle in cases suspected of dilated cerebral ventricles in late gestation.

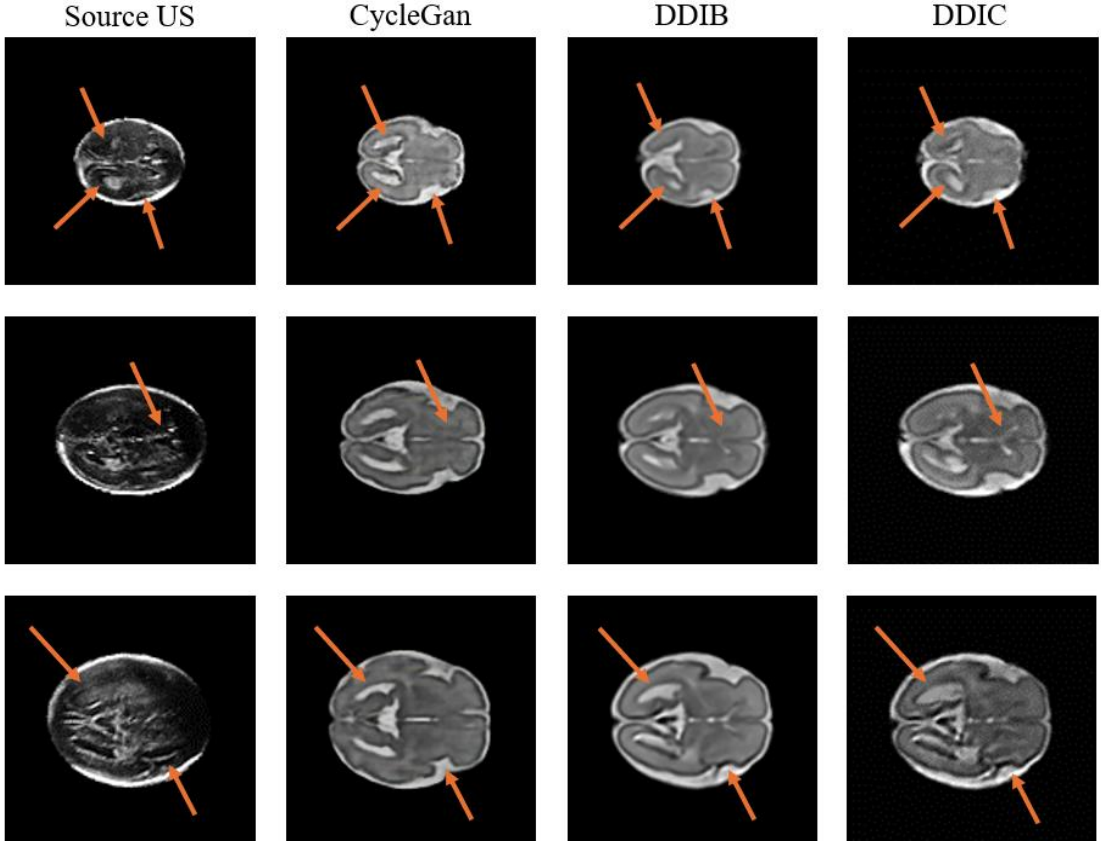

Figure 3: Qualitative evaluation of US to pseudo-MRI translation. Each row presents an exemplary sample from a different gestational age. (1st column) original US image with head segmentation. (2nd column) CycleGAN pseudo-MRI synthesis. (3rd column) DDIB pseudo-MRI synthesis. (4th column) DDIC pseudo-MRI synthesis. The arrows indicate zones of inconsistency between the pseudo-MRI reconstructions and the original US. Note the improved performance of the DDIC.

In comparison to CycleGAN and DDIB, the DDIC successfully preserves fine details between the original US image and its translated counterpart, as demonstrated in Figure 3. The arrows delineate regions wherein the DDIC algorithm exhibits enhancements in illustrating fine components of the fetal brain structures. In the first row, the bottom right arrow highlights the insula region, wherein CycleGAN and DDIB methods exhibit distortion in the sulcation. In contrast, the DDIC method preserves the original US scan insula structure. The two other arrows indicate the lateral ventricles. CycleGAN failed to maintain the structure of the lateral ventricles, leading to their enlargement and elongation, wrongly indicative of a more advanced gestational stage compared to the original ultrasound image. Similarly, DDIB failed to preserve the distinctiveness of the lateral ventricles, amalgamating them with the flax midline. In contrast, the DDIC algorithm successfully separated the lateral ventricles from the flax midline and preserved their original anatomical structure. In the second row, the arrow denotes the Cavum Septum Pellucidum (CSP). As can be observed, the DDIC algorithm exhibits the highest level of sharpness and distinction. The top arrow in the third row indicates the left lateral ventricle. DDIC preserves the size and shape of the ventricle as depicted in the US image, as opposed to the other methods. The lower arrow signifies the maintenance of the insula sulcation, consistent with the depiction provided by the images in the first row.

The quantitative assessment is graphically depicted in Figure 4, wherein the evaluation metrics of MI, PSNR, and FID are presented. As can be observed, the DDIC yields the best results in all categories. These findings indicate that the performance of the proposed method for translating US to pseudo-MRI surpasses that of competing architectures.

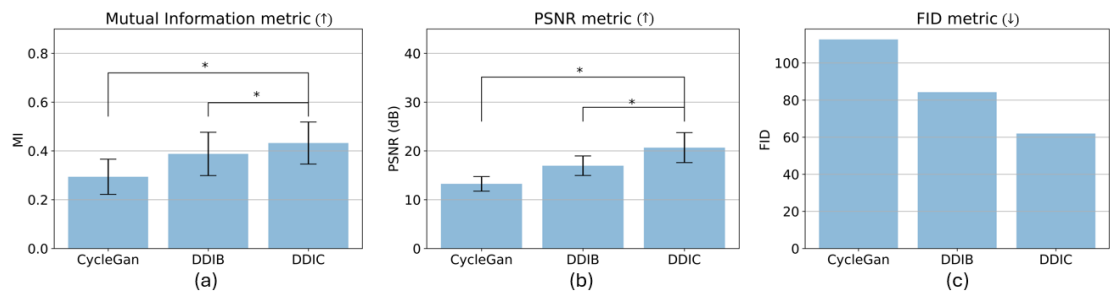

Figure 4: (a) Distributions of Mutual Information metric; (b) Distributions of PSNR metric; (c) Distributions of FID metric. *Indicates statistically significant two-sided P value < 0.001 with the unequal variances t-test.

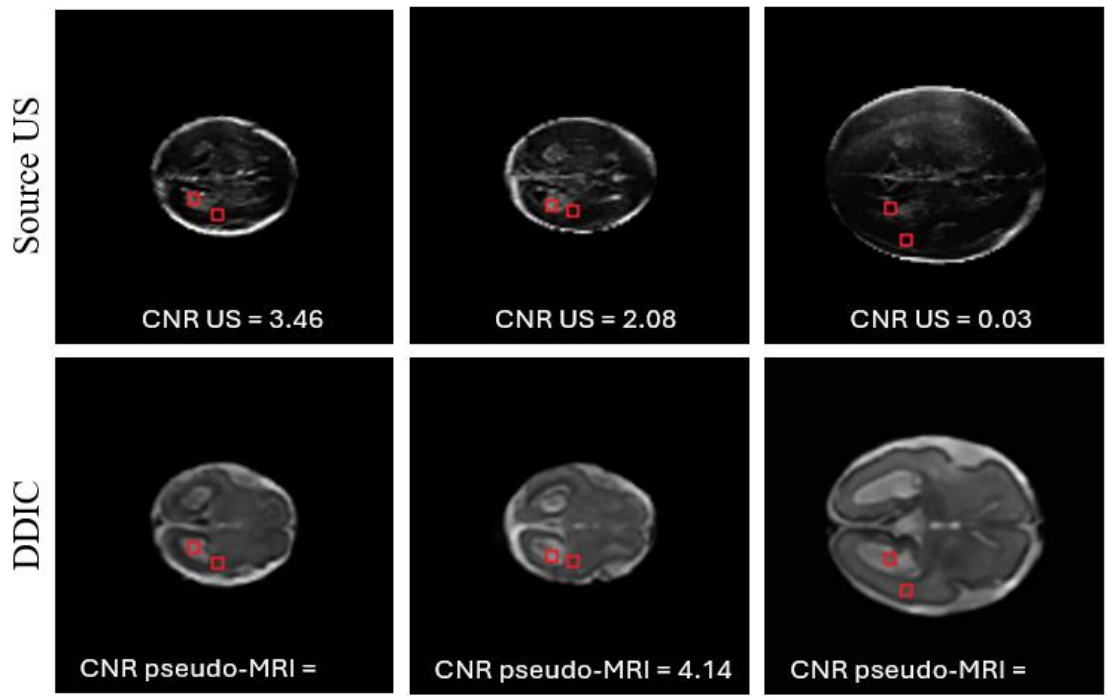

Figure 5: Three examples of the CNR zones selection. The US image is on the left, and the DDIC pseudo-MRI is on the right. The upper square on the lateral ventricle is the region of interest (ROI), and the lower square is the background for the CNR calculation.

It is noteworthy to mention our attempt to implement the methodology suggested by Jiao *et al.* (7), which translates fetal brain US to pseudo-MRI with a GAN-based model. However, we obtained meaningless pseudo-MRI images. We postulate that this outcome may be attributed to the comparatively limited size of the datasets available in our study in contrast to the extensive datasets employed by Jiao *et al.*

To demonstrate the enhancement in contrast of the pseudo-MRI images relative to the original ultrasound images, we computed the CNR as outlined in the method section. Representative contrast regions are illustrated in Figure 5. The results indicate that the contrast in the DDIC pseudo-MRI images is improved, from $1.37 \pm 1.24$ in the US images to $2.61 \pm 1.75$ in the pseudo-MRI images. The observed standard deviation can be attributed to variations in gestational age.

To illustrate the visual advantages of the DDIC-generated pseudo-MRI images compared to the original US images, we applied the "Segment Anything" algorithm on the test set images as described in the methods section. The pseudo-MRI images consistently yielded more reliable segmentation than both the full-resolution and low-resolution US images. Exemplary results are shown in Figure 7. In each of the presented examples, the segmentation of the pseudo-MRI images, compared to the US images, is noticeably smoother, free of irregularities, and successfully captures the entire target surface, including the contours of the lateral ventricles and the Sylvian fissure.

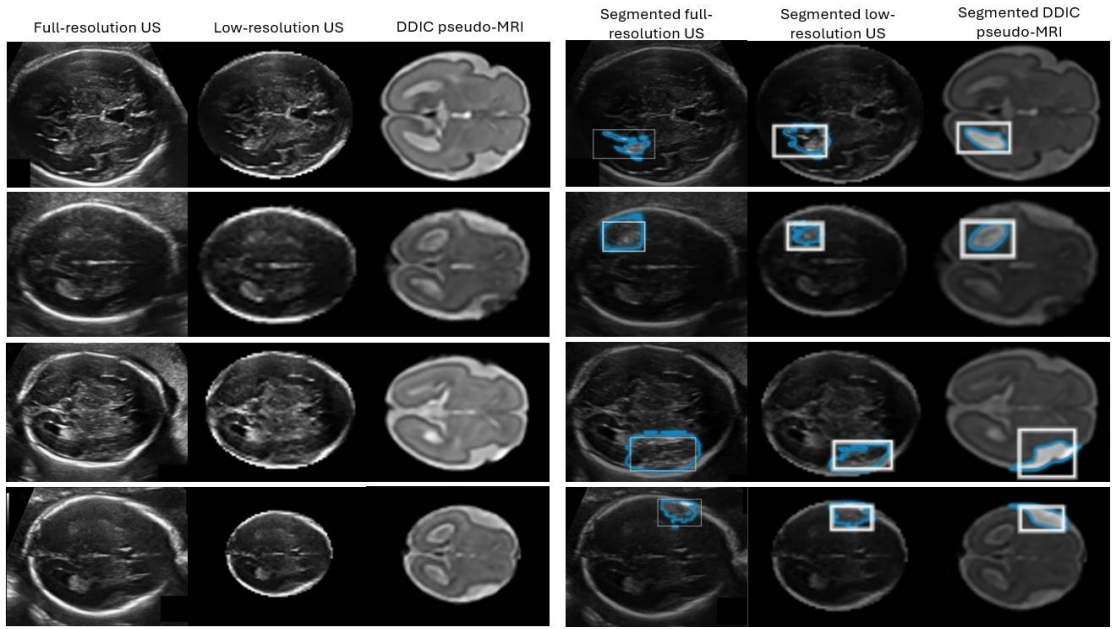

Figure 6: An example of Meta AI's "Segment Anything" algorithm demonstrates its application in segmenting anatomical features on three types of images: The original full-resolution US image as taken from the HC18 dataset, a low-resolution US image as preprocessed for data input to the DDIC algorithm, and the DDIC pseudo-MRI image. (1st row) Distal lateral ventricle segmentation. (2nd row) Proximal lateral ventricle segmentation. (3rd row) Distal Sylvian fissure segmentation, (4th row) Proximal Sylvian fissure segmentation. As can be observed, the segmentations in the pseudo-MRI images are noticeably smoother and more distinct compared to the segmentations of the original US images (both full and low resolution).

To validate the clinical value of the image translation, we conducted the MOT test as described in the methods section. The results are graphically presented in Figure 6. The findings reveal that, on average, the physicians reported an improvement in clinical information in approximately 81% of the evaluated images (see Figure 6a). The most significant improvement occurred in the proximal Sylvian fissure. Notably, there is a more significant enhancement in the proximal area, particularly concerning the lateral ventricle and the Sylvian fissure (see Figure 6b). These regions, typically not visible in ultrasound images (31), show marked improvement, potentially facilitating more accurate diagnoses.

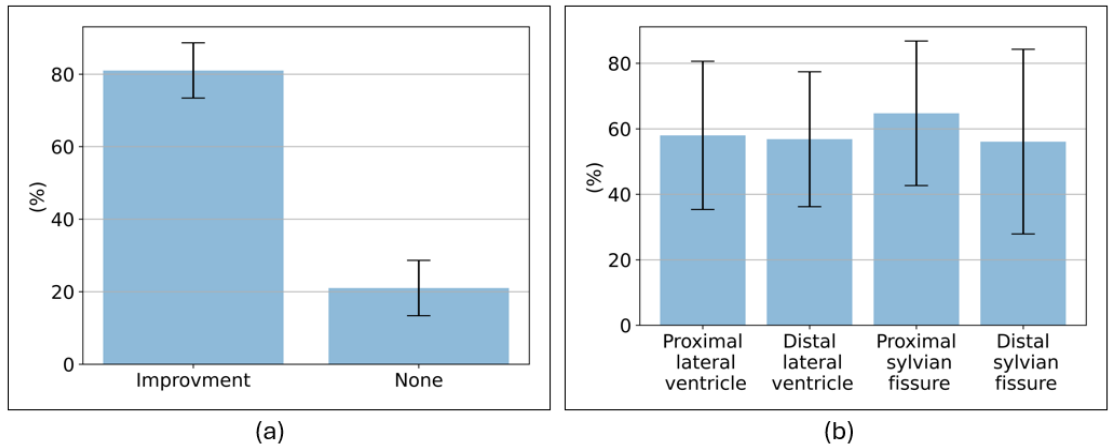

Figure 7: Graphical depiction of the MOT score. (a) Percentage of images in which the physicians have found improvement in at least one feature. (b) The percentage of images from the improved group in which each specific feature was improved. Notably, there is a significant enhancement observed in all features relative to the original US.

## 4  Discussions

Ultrasonic fetal images pose a challenge due to their inherent low image quality which is characterized by acoustic shadows and speckled texture. These obstructive factors are more prominent in the proximal zone and may be crucial when suspected anomalies may be present. In this study, we have demonstrated that the clinical merit of the acquired US images can be augmented by implementing image translation into pseudo-MRI display using the DDIC method. As observed, 81% of the tested translated images were improved according to the MOT test.

The primary advantage of the DDIC method lies in the improved visibility of the lateral ventricles and the Sylvian fissure (as shown in Figure 6b), potentially leading to a better fetal development diagnosis. This is achieved by the integrated optimization which imposes the correlation between the resultant pseudo-MRI image and the original ultrasound image. This correlation facilitates similarity in the fine details within the brain structure, leading to a more straightforward clinical interpretation.

Furthermore, as can be observed from the results, DDIC has improved image quality in all the quantitative metrics studied here (Figs. 3, 4, 5). Notably, the CNR measured for the distal ventricle has almost doubled in the generated images compared to the ultrasonic source.

Importantly, the Mutual Information (MI) metric was better relative to the reference translation methods applied in this study. The MI metric for translation quality evaluation allows for measuring the structural similarity between the US and MRI images without necessitating signal similarity, thereby facilitating direct comparison. This is very important since MRI and US have opposing depictions of certain anatomical features, for example, the ventricles appear black in the US but white in the MRI. In addition, the Fréchet inception distance (FID) which is a metric used to assess the quality of images created by a generative model, was also superior (smaller).

The DDIC algorithm offers a notable advancement in the application of segmentation algorithms for identifying brain anatomy, demonstrating superior performance on synthesized pseudo-MRI images compared to the source ultrasound images. In particular, identifying the lateral ventricles and the Sylvian fissure which are more distinctly visible in the pseudo-MRI images, exhibiting enhanced contrast and sharper boundaries, as confirmed by the MOT test. Figure 7 provides examples of the "Segment Anything" algorithm developed by Meta AI (30), which achieves significantly improved segmentations of the lateral ventricles and the Sylvian fissure in the pseudo-MRI images compared to the ultrasound images. These segmentations hold significant clinical value. As noted in the introduction, identifying the shape of the Sylvian fissure and evaluating the size of the lateral ventricles are crucial for diagnosing cortical dysplasia and fetal cerebral lateral ventriculomegaly respectively (17,31) We have demonstrated the application of the algorithm on both the original full-resolution ultrasound image and the low-resolution preprocessed ultrasound image to highlight the advantage of the DDIC even though it reduces the original resolution of the ultrasound images.

Another advantage of the DDIC method stems from the fact that it performs separate training for the ultrasound and MRI image databases. This separated training approach offers the potential to utilize databases taken from various sources. It enables training the US model in one hospital and the MRI model in another, eliminating the need for patient images to leave hospital premises and maintaining patient confidentiality. Moreover, in contrast to GAN-based models, employing a diffusion model for image translation enables the learning process to generate new images effectively even with a relatively small dataset. This capability represents a significant advantage, especially in the medical domain where access to extensive databases is often restricted.

Another point worth mentioning is the limited availability of databases containing fetal MRI images. This stems from the fact that unlike US, fetal MRI examinations are not routinely preformed. Hence presumably, US-MRI translation presents an opportunity to generate fetal pseudo-MRI datasets from widely accessible ultrasound databases. This process can facilitate the training of MRI-based models for various medical applications. Moreover, image translation can aid in the development of registration algorithms between US and MRI modalities (32). Such algorithms necessitate the identification of salient points present in both images. When the images are displayed in the same format, the process of identifying these points becomes more facile.

# 5 Conclusion

This study introduces a technique for translating fetal US images into a pseudo-MRI display, aimed at improving the visualization of fetal brain anatomical structures. Notably, the resulting pseudo-MRI Sylvian fissure, characterized by well-defined borders and significantly improved contrast clarity according to the MOT. This enhancement holds promise for expediting and enhancing the accuracy of fetal growth diagnosis, potentially leading to improved clinical outcomes.

**Acknowledgments**

This work was supported in part by the Ministry of Innovation, Science and Technology, Israel, under Grant #880011 and in part by the Technion Israel Institute of Technology scholarship.

24. Gholipour A, Limperopoulos C, Clancy S, Clouchoux C, Akhondi-Asl A, Estroff JA, et al. LNCS 8674 - Construction of a Deformable Spatiotemporal MRI Atlas of the Fetal Brain: Evaluation of Similarity Metrics and Deformation Models. 2014.

25. Payette K, de Dumast P, Kebiri H, Ezhov I, Paetzold JC, Shit S, et al. An automatic multi-tissue human fetal brain segmentation benchmark using the Fetal Tissue Annotation Dataset. Sci Data. 2021 Dec 1;8(1).

26. Jiao J, Namburete AIL, Papageorghiou AT, Noble JA. Self-Supervised Ultrasound to MRI Fetal Brain Image Synthesis. IEEE Trans Med Imaging. 2020 Dec 1;39(12):4413–24.

27. Kraskov A, Stögbauer H, Grassberger P. Estimating mutual information. Phys Rev E Stat Phys Plasmas Fluids Relat Interdiscip Topics. 2004;69(6):16.

28. Horé A, Ziou D. Image quality metrics: PSNR vs. SSIM. In: Proceedings - International Conference on Pattern Recognition. 2010. p. 2366–9.

29. Yu Y, Zhang W, Deng Y. Frechet Inception Distance (FID) for Evaluating GANs Beijing, China:China Univ. Mining Technol. Beijing Graduate School, 2021.

30. Kirillov A, Mintun E, Ravi N, Mao H, Rolland C, Gustafson L, et al. Segment Anything. 2023 Apr 5; Available from: http://arxiv.org/abs/2304.02643

31. Ganor Paz Y, Levinsky D, Rosen H, Barzilay E. Feasibility of Fetal Proximal Lateral Cerebral Ventricle Measurement. Journal of Ultrasound in Medicine. 2022 Dec 1;41(12):2933–8.

32. Fuerst B, Wein W, Müller M, Navab N. Automatic ultrasound-MRI registration for neurosurgery using the 2D and 3D LC2 Metric. Med Image Anal. 2014;18(8):1312–9.
13